\documentclass[12pt]{iopart}
\usepackage{amssymb}
\usepackage{setstack}
\usepackage{graphicx,bm}

\begin{document}

\title[Subsurface Kondo impurities]{The signature of subsurface Kondo
impurities in the local tunnel current.}
\author{Ye.S. Avotina $^{1,2}$, Yu.A. Kolesnichenko $^{1,2}$, and J.M. van
Ruitenbeek $^{2}$}

\begin{abstract}
The conductance of a tunnel point-contact in an STM-like geometry having a
single defect placed below the surface is investigated theoretically. The
effect of multiple electron scattering by the defect after reflections by
the metal surface is taken into account. In the approximation of s-wave
scattering the dependence of the conductance on the applied voltage and the
position of the defect is obtained. The results are illustrated for a model
s-wave phase shift describing Kondo-resonance scattering. We demonstrate
that multiple electron scattering by the magnetic impurity plays a decisive
role in the point-contact conductance at voltages near the Kondo resonance.
We find that the sign and shape of the Kondo anomaly depends on the position
of the defect.
\end{abstract}


\address{$^{1}$ B.I. Verkin Institute for Low Temperature
Physics and Engineering, National Academy of Sciences of Ukraine,
47, Lenin Ave., 61103, Kharkov,Ukraine.}
\address {$^{2}$
Kamerlingh Onnes Laboratorium, Universiteit Leiden, Postbus 9504,
2300 Leiden, The Netherlands.}

\eads{kolesnichenko@ilt.kharkov.ua}

\submitto{JPCM}
\maketitle
\section{Introduction}

Various surface defects have been observed and investigated by scanning
tunneling microscopy (STM) \cite{Crommie,Ber,Koles,Knorr}. The interference
of the surface electron waves results in an oscillatory dependence of the
tunneling conductance measured as a function of the separation between the
STM tip and the defect. Remarkable manifestations of quantum interference
were observed in artificial structures built from single atoms on a clean
metal surface, the so-called quantum corrals \cite{corral}. Magnetic adatoms
on non-magnetic host metal surfaces are of special interest as they produce
a characteristic many-body resonance structure in the differential
conductance near zero voltage bias attributed to the Kondo effect \cite%
{chen99,madhavan98,li98,knorr02}. The surface electrons waves contain the
information of the magnetic impurity and by focussing the waves it has been
possible to create a mirage image of the impurity \cite{manoharan00}. The
shape of the resonance in the differential conductance, $dI/dV$, is usually
asymmetric and is described by a Fano line shape \cite%
{fano61,plihal01,Zawad,wahl04}.

In principle STM spectroscopy should also provide access to information on
the structure of the metal \emph{below} the surface. This possibility is
based on the influence on the conductance caused by quantum interference of
electron waves that are scattered by defects and reflected back by the
contact. This effect was explored by Schmidt and coworkers \cite{babble} for
investigating subsurface bubbles of implanted gas in Al. The observation of
interference patterns due to electron scattering by Co impurities in the
interior of a Cu sample was reported by Quaas et al. \cite{Wend}.
Theoretically, the influence of single defects in the bulk of a metal on the
quantum conductance of tunnel point-contact has been discussed in Refs.~\cite%
{Avotina1,Avotina2,Avotina3}. In these papers it has been shown that the
location of defects below the surface can be identified from the
interference pattern in constant-current STM images combined with the
information obtained from the dependence of the conductance on the applied
voltage. In the previous work of Refs.~\cite{Avotina1,Avotina2,Avotina3} the
scattering of electrons with a defect has been taken into account in the
framework of perturbation theory. Such an approximation is valid as long as
the strength of the electron - impurity scattering interaction is small. In
the case of a magnetic defect at low temperatures ($T\ll T_{K}$, where $%
T_{K} $ is the Kondo temperature) the Kondo resonance results in a dramatic
enhancement of the effective electron-impurity interaction \cite{Abrikosov}
and the perturbation method becomes inapplicable.

In this paper we present the quantum conductance $G$ of the tunnel point
contact in the vicinity of which a single point-like defect is situated, for
arbitrary values of the scattering potential. We express the conductance by
the means of a s-wave scattering phase shift $\delta _{0}$. The results
describe the influence to the conductance of multiple scattering of the
electrons by a single defect. Multiple scattering needs to be included even
for a single defect because of electron reflection by the metal surface.
This results in the appearance of harmonics in the dependence of $G$ on the
applied voltage and on the distance between the contact and the defect. We
apply the analysis of the non-monotonic voltage dependence of the
conductance specifically for the interesting problem of Kondo scattering,
using an appropriate phase shift \cite{Shift}. To our knowledge, observation
of subsurface Kondo impurities have not yet been reported in experiments,
and the present analysis may guide future experimental investigations.

\section{Model and basic equations}

\begin{figure}[tbp]
\includegraphics [width=8cm]{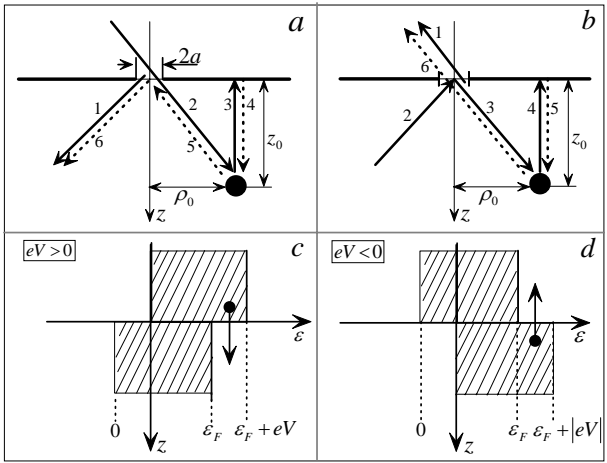}
\caption{$(a,b)$ Model of the contact and $(c,d)$ illustration of the
occupied energy bands in the two metal half-spaces for both signs of the
applied bias $eV$. In panels $(a,b)$ the defect is placed at the point $%
\mathbf{r}_{0}=\left( \mathbf{\protect\rho }_{0},z_{0}\right)$. Electron
trajectories are shown schematically. Note that we take the $z$-axis
pointing downward.}
\label{fig1}
\end{figure}

In our model of the system we represent the contact by an orifice of radius $%
a$ centered at the location of the 'STM tip', $\mathbf{r}=0$. The orifice
provides a tunneling window in otherwise impenetrable infinitely thin
interface at $z=0$ between two metal half-spaces (Fig.~\ref{fig1}). The
potential barrier at the plane of interface, $z=0$, is taken to be described
by a delta function, $U\left( \mathbf{r}\right) =U_{0}f\left( \rho \right)
\delta \left( z\right) ,$ where $\rho $ is the length of the radius vector $%
\mathbf{\rho }$ in the plane $z=0$. The function $f\left( \rho \right)
\rightarrow \infty $ in all points of the plane except in the contact, where
$f\left( \rho \right) =1$. At the point $\mathbf{r}_{0}$ a defect described
by the potential $D\left( \left\vert \mathbf{r-r}_{0}\right\vert \right) $
is placed.

We consider an almost ballistic configuration (the electrons are
elastic scattered by the single defect only) and neglect
electron-phonon scattering assuming the electron mean free path to
be much large than the distance between the contact and the
defect. In Ref. \cite{Agrait} the authors reported the observation
of conductance oscillations at a voltage range up to 1.5$eV$ at a
temperature of 4.2K. Large bias voltages can be applied to small
tunnel junctions created by STM or break-junction methods without
significant heating of the electrodes. Because of the high
resistance of the contact the
current density remains small. Below we restrict our plots by the range $%
eV<\varepsilon _{F}$.

For the host metal we will consider a free electron model with an electron
effective mass $m^{\ast }$ and a dispersion relation $\varepsilon _{\mathbf{k%
}}=\hbar ^{2}k^{2}/2m^{\ast }$, where $\mathbf{k}$, and $\varepsilon _{%
\mathbf{k}}$ are the electron wave vector and electron energy, respectively.
The electron wave function $\psi _{\mathbf{k}}$ satisfies the Schr\"{o}%
dinger equation
\begin{equation}
\eqalign{\frac{\hbar ^{2}}{2m^{\ast }}\nabla ^{2}\psi _{\mathbf{k}}\left( \mathbf{r}%
\right) +\left[ \varepsilon _{\mathbf{k}}-U\left(
\mathbf{r}\right) -V\left( \mathbf{r}\right) \right] \psi
_{\mathbf{k}}\left( \mathbf{r}\right) =
\label{Schrod} \\
D\left( \left\vert \mathbf{r-r}_{0}\right\vert \right) \psi _{\mathbf{k}%
}\left( \mathbf{r}\right) ,}
\end{equation}%
where the $V\left( \mathbf{r}\right) $ is the applied electrostatic
potential. The function $\psi _{\mathbf{k}}\left( \mathbf{\rho },z\right) $
satisfies boundaries conditions of continuity and of the jump of its
derivative at the boundary $z=0.$ We will assume that the transmission
amplitude of electrons through the barrier in the orifice is small,
\begin{equation}
t\left( k\right) \approx \hbar ^{2}k/im^{\ast }U_{0};|t|\ll 1.  \label{t}
\end{equation}%
For small transparency $t$ the voltage drop due to the applied bias is
entirely localized at the barrier. The electric potential can be described
by a step function, $V\left( z\right) =V\,\Theta \left( -z\right) .$ As a
result, the occupied energy bands in the half-spaces $z>0$ and $z<0$ are
shifted by $eV$. We take the zero of energy, $\varepsilon =0$, to coincide
with the bottom of the lower of the two bands, i.e. $\varepsilon =0$ lies at
the bottom of the band in the half-space $z>0$ when $eV>0$ and at the bottom
of the band in the half-space $z<0$ for $eV<0.$ At zero temperature
electrons tunnel to the lower half-space (Fig.~\ref{fig1}(c,d)) when $eV>0$,
and for $eV<0$ electrons can tunnel only to available states in the upper
half-space.

As shown in Refs.~\cite{KMO,Avotina1} Eq.~(\ref{Schrod}) can be solved for
arbitrary form of the function $f\left( \rho \right) $ in the limit $%
\left\vert t\right\vert \rightarrow 0.$ To first approximation in the small
parameter $|t|\ll 1$ (\ref{t}) the wave function $\psi _{\mathbf{k} }\left(
\mathbf{r}\right) $ can be written as:
\begin{equation}
\psi _{\mathbf{k}}\left( \mathbf{r}\right) =\psi _{\mathbf{k}0}\left(
\mathbf{r}\right) +\psi _{\mathbf{k}1}\left( \mathbf{r}\right) ,
\label{_psi}
\end{equation}%
where $\psi _{\mathbf{k}1}\left( \mathbf{r}\right) \sim 1/U_{0}.$ This
latter part of the wave function (\ref{_psi}) describes the electron
tunnelling through the barrier and determines the electrical current. The
first term in the Eq.~(\ref{_psi}) is the solution of the Schr\"{o}dinger
equation for the metallic half-spaces without the contact. It satisfies the
boundary condition $\psi _{\mathbf{k}0}\left( \mathbf{\rho },0\right) =0$ at
the interface.

For $|t|\ll1$ the boundary condition for the jump of the derivative of the
total wave function is reduced to \cite{KMO}
\begin{equation}
\mp \left. \frac{\partial \psi _{\widetilde{\mathbf{k}}0}^{\left( \mp
\right) }}{\partial z}\right\vert _{z=\mp 0}=\frac{2m^{\ast }}{\hbar ^{2}}%
U_{0}f\left( \rho \right) \psi _{\mathbf{k}1}^{\left( \pm \right) }\left(
\mathbf{\rho },0\right) ,
\end{equation}%
where $\psi _{\mathbf{k}s}^{\left( \pm \right) }$ $\left( s=0,1\right) $ are
the wave functions for $z\gtrless 0$, $\widetilde{\mathbf{k}}$ is the
electron wave vector for electrons arriving in one half-space from the
another half-space through the orifice, $(|\widetilde{\mathbf{k}}|=\sqrt{%
k^{2}-2m^{\ast }\left\vert eV\right\vert /\hbar ^{2}})$.

Thus, the function $\psi _{\mathbf{k}1}\left( \mathbf{r}\right) $ can be
expressed by means of the solution $\psi _{\mathbf{k}0}\left( \mathbf{r}%
\right) .$ By using the Fourier transform of the wave function (\ref{_psi})
we find%
\begin{eqnarray}
\psi _{\mathbf{k}1}^{\left( \pm \right) }\left( \mathbf{r}\right)  &=&\mp
\frac{\hbar ^{2}}{2m^{\ast }U_{0}}\int\limits_{-\infty }^{\infty }d\mathbf{%
\kappa }^{\prime }e^{i\mathbf{\kappa }^{\prime }\mathbf{\rho +}%
ik_{z}^{\prime }\left\vert z\right\vert }\left. \frac{\partial \psi _{%
\widetilde{\mathbf{k}}0}^{\left( \mp \right) }}{\partial z}\right\vert
_{z=\mp 0}\times  \\
&&\frac{1}{\left( 2\pi \right) ^{2}}\int\limits_{-\infty }^{\infty }d\mathbf{%
\rho }^{\prime }\frac{e^{i\mathbf{\kappa }^{\prime }\mathbf{\rho }^{\prime }}%
}{f\left( \rho \right) },  \nonumber
\end{eqnarray}%
where $k_{z}^{\prime }=\sqrt{k^{2}-\kappa ^{\prime 2}}.$ The electron wave
function, $\psi _{\mathbf{k}}\left( \mathbf{r}\right) $, \ which takes into
account \ the scattering by the defect, can be expressed by means of the
retarded Green function $G_{0}^{+}\left( \mathbf{r}^{\prime }\mathbf{,r}%
;\varepsilon \right) $ of the homogeneous equation (\ref{Schrod}) at $D=0$
and $U\rightarrow \infty .$ To first approximation in the transmission
amplitude $t$ (\ref{t}) the integral equation for $\psi _{\mathbf{k}s}\left(
\mathbf{r}\right) $ is given by
\begin{equation}
\eqalign{\psi _{\mathbf{k}s}\left( \mathbf{r}\right) =\psi
_{\mathbf{k}s}^{\left(
0\right) }\left( \mathbf{r}\right) +  \label{phi} \\
\frac{2m^{\ast }}{\hbar ^{2}}\int d\mathbf{r}^{\prime }D\left( \left\vert
\mathbf{r}^{\prime }-\mathbf{r}_{0}\right\vert \right) G_{0}^{+}\left(
\mathbf{r,r}^{\prime };\varepsilon \right) \psi _{\mathbf{k}s}\left( \mathbf{%
r}^{\prime }\right) ,}
\end{equation}%
where
\begin{equation}
G_{0}^{+}\left( \mathbf{r,r}^{\prime };\varepsilon \right) =-\frac{ik}{4\pi }%
\left\{ h_{0}^{\left( 1\right) }\left( k\left\vert \mathbf{r}-\mathbf{r}%
^{\prime }\right\vert \right) -h_{0}^{\left( 1\right) }\left( k\left\vert
\mathbf{r}-\widetilde{\mathbf{r}}^{\prime }\right\vert \right) \right\} ,
\label{G+0}
\end{equation}%
\ $\widetilde{\mathbf{r}}^{\prime }=\left( \mathbf{\rho }^{\prime
},-z^{\prime }\right) .$ In Eq.~(\ref{G+0}) and below $h_{l}^{\left(
1\right) }\left( x\right) $ are the spherical Hankel functions. The first
term in the braces is the Green function for free electrons in the infinite
space and the second one takes into account the specular electron reflection
from the interface. The functions $\psi _{\mathbf{k}s}^{\left( 0\right)
}\left( \mathbf{r}\right) $ are the wave functions to zeroth and first order
in $t$ in the absence of the defect ($D=0$). The electron wave function in
the metal half-spaces is
\begin{equation}
\psi _{\mathbf{k}0}^{\left( 0\right) }\left( \mathbf{r}\right) =e^{i\mathbf{%
\kappa \rho }}\left( e^{ik_{z}\left\vert z\right\vert }-e^{-ik_{z}\left\vert
z\right\vert }\right) ,  \label{psi00}
\end{equation}%
where $\mathbf{\kappa }$ and $k_{z}$ are the components of the vector $%
\mathbf{k}$ parallel \ and perpendicular to the interface. The wave function
$\psi _{\mathbf{k}1}^{\left( 0\right) }\left( \mathbf{r}\right) $ of the
electrons that are transmitted through the contact has been obtained in Ref.~%
\cite{KMO}. In order to simplify further calculations we consider a point
contact, taking the limit $a\rightarrow 0$. The solution $\psi _{\mathbf{k}%
1}^{\left( 0\right) }\left( \mathbf{r}\right) $ in this limit is given in
Ref.~\cite{Avotina2} for any arbitrary anisotropic quadratic electron
dispersion law $\varepsilon _{\mathbf{k}}$. For an isotropic band $%
\varepsilon _{\mathbf{k}}=\hbar ^{2}k^{2}/2m^{\ast }$ this takes the form,
\begin{equation}
\psi _{\mathbf{k}1}^{\left( 0\right) }\left( \mathbf{r}\right) =t\left(
\widetilde{k}_{z}\right) \frac{i\left( ka\right) ^{2}\cos \theta }{2}%
h_{1}^{\left( 1\right) }\left( kr\right) .  \label{ps01}
\end{equation}%
Here, $\left( r,\theta ,\varphi \right) $ are the spherical coordinates of
the vector $\mathbf{r}$, with $\theta $ the angle between the vector $%
\mathbf{r}$ and the contact axis. $\widetilde{k}_{z}$ is the $z-$component
of the vector $\widetilde{\mathbf{k}}$ . The plane wave (\ref{psi00}) is
transformed into a spherical p-wave $h_{1}^{\left( 1\right) }\left(
kr\right) $ (\ref{ps01}) after scattering by the point contact.

This model allows us to solve the three dimensional
Schr\"{o}dinger equation in the limit of small transparency of the
barrier and find the analytical formulas for the conductance. Our
method is similar to the widely employed tunneling Hamiltonian
approach, where in the limit of small transparency of the barrier
the distribution functions of electrons in the electrodes can be
taken to be in equilibrium (Fermi functions) with chemical
potentials shifted by the bias $eV$. For a barrier of finite width
the electric field distribution changes which influences the
nonlinear dependence of the conductance. This dependence becomes
very important if the bias is comparable with the work function of
the metal. For any three dimensional models of the potential
barrier the dependence $G\left( V\right) $ may be calculated only
numerically. In our paper we did not make it our aim to
investigate the intrinsic conductance of the tunnel junction
$G_{0}\left( V\right) $. The purpose of the work is to investigate
the oscillatory and resonance additions to the conductance
$G_{0}\left( V\right) $ in the presence of a defect in the bulk of
the metal, where the distribution functions are in equilibrium (in
leading approximation in the barrier transparency). We believe
that the part of the conductance related to
the defect, which will be obtained in next sections, is correct, if the bias $eV$ is less than Fermi energy $%
\varepsilon _{F}$.

\section{Scattered wave function in s-wave approximation}

Let $D\left( \left\vert \mathbf{r}-\mathbf{r}_{0}\right\vert \right) $ be a
spherically symmetric scattering potential which is finite in the point $%
\mathbf{r}=\mathbf{r}_{0}$ and tends to zero at a distance $r_{D}\ll r_{0}$
that is of the order of the Fermi wave length $\lambda _{\mathrm{F}}$. As is
well known, s-wave scattering is dominant for scattering by a short range
potential \cite{Davidov}. In order to express the wave function (\ref{phi})
by the s-wave phase shift $\delta _{0}$ we use the 'sharpness' of the
function $D\left( \left\vert \mathbf{r}^{\prime }-\mathbf{r} _{0}\right\vert
\right)$, which essentially differs from zero only in a small region of the
radius $r_{D}$ near the point $\mathbf{r}^{\prime }= \mathbf{r}_{0}.$ The
main contribution to the integral in Eq.~(\ref{phi}) comes from this region
and the 'smooth' functions $\psi _{\mathbf{k}s}\left( \mathbf{r}^{\prime
}\right) $ and $h_{0}^{\left( 1\right) }\left( k\left\vert \mathbf{r}-%
\widetilde{\mathbf{r}}^{\prime }\right\vert \right) $ can be taken outside
the integral at the point $\mathbf{r}^{\prime }= \mathbf{r}_{0}$. For $%
\left\vert \mathbf{r}-\mathbf{r}_{0}\right\vert \gg r_{D}$ the solution of
Eq.~(\ref{phi}) takes the form \cite{Avotina2},
\begin{equation}
\psi _{\mathbf{k}s}\left( \mathbf{r}\right) \approx \psi _{\mathbf{k}%
s}^{\left( 0\right) }\left( \mathbf{r}\right) +\frac{2m^{\ast }}{\hbar ^{2}}%
T\left( k\right) \psi _{\mathbf{k}s}^{\left( 0\right) }\left( \mathbf{r}%
_{0}\right) G_{0}^{+}\left( \mathbf{r,r}_{0};\varepsilon \right) ,
\label{psi_scat}
\end{equation}%
where%
\begin{equation}
T\left( k\right) =\frac{g}{1+\frac{m^{\ast }ik}{2\pi \hbar ^{2}}\left[
Y\left( k\right) -gh_{0}^{\left( 1\right) }\left( 2kz_{0}\right) \right] },
\label{T}
\end{equation}%
\begin{equation}
Y\left( k\right) =\int d\mathbf{r}^{\prime }D\left( r^{\prime }\right)
h_{0}^{\left( 1\right) }\left( kr^{\prime }\right), g=\int d\mathbf{r}%
^{\prime }D\left( r^{\prime }\right) .  \label{J_p}
\end{equation}

Let us compare the wave function (\ref{psi_scat}) with the formal solution $%
\psi _{\mathbf{k}}^{sc}\left( \mathbf{r}\right) $ of the scattering problem
for the spherically symmetrical potential $D\left( \left\vert \mathbf{r}-%
\mathbf{r}_{0}\right\vert \right) $ in infinite space
\begin{equation}
\psi _{\mathbf{k}}^{sc}\left( \mathbf{r}\right) \approx \psi _{\mathbf{k}%
}^{in}\left( \mathbf{r}\right) -\frac{im^{\ast }k}{2\pi \hbar ^{2}}%
T_{0}\left( k\right) \psi _{\mathbf{k}}^{in}\left( \mathbf{r}_{0}\right)
h_{0}^{\left( 1\right) }\left( k\left\vert \mathbf{r}_{0}-\mathbf{r}%
\right\vert \right) ,  \nonumber
\end{equation}%
where $\psi _{\mathbf{k}}^{in}$ and$\ \psi _{\mathbf{k}}^{sc}$ are incident
and scattered waves, and
\begin{equation}
T_{0}\left( k\right) =\frac{g}{1+\frac{m^{\ast }i k}{2\pi \hbar ^{2}}Y\left(
k\right) },
\end{equation}%
\ \ is the $T$ matrix. Taking into account the relation between $T_{0}$ and
the s-wave phase shift $\delta _{0}\left( k\right) $
\begin{equation}
-\frac{m^{\ast }}{2\pi \hbar ^{2}}T_{0}=\frac{1}{k}e^{i\delta _{0}}\sin
\delta _{0},
\end{equation}%
we rewrite the Eq.~(\ref{T}) in the form
\begin{equation}
T\left( k\right) =-\frac{\pi \hbar ^{2}}{m^{\ast }ik}\frac{e^{2i\delta
_{0}}-1}{1+\frac{1}{2}\left( e^{2i\delta _{0}}-1\right) h_{0}^{\left(
1\right) }\left( 2kz_{0}\right) }.  \label{T_d}
\end{equation}%
Note that the effective $T$-matrix (\ref{T_d}) is an oscillatory function of
the distance $z_{0}$ between the defect and the interface that results from
repeated electron scattering by the defect after its reflections from the
interface.

For a calculation of the current we should know the wave functions of the
electrons transmitted through the contact, from one half-space to the other.
For $z>0$ and $eV>0$ (i.e. for electron tunneling into the half-space in
which the defect is situated) we find%
\begin{equation}
\eqalign{\psi _{\mathbf{k}1}^{\left( +\right) }\left( \mathbf{r}\right) =\psi _{%
\mathbf{k}1}^{\left( 0\right) }\left( \mathbf{r}\right) -  \label{psi111} \\
\frac{m^{\ast }ik}{2\pi \hbar ^{2}}T\left( k\right) \psi _{\mathbf{k}%
1}^{\left( 0\right) }\left( \mathbf{r}_{0}\right) \left\{ h_{0}^{\left(
1\right) }\left( k\left\vert \mathbf{r}-\mathbf{r}_{0}\right\vert \right)
-h_{0}^{\left( 1\right) }\left( k\left\vert \mathbf{r}-\widetilde{\mathbf{r}}%
_{0}\right\vert \right) \right\} .}
\end{equation}%
For $z<0$ and $eV<0$ (i.e. for electron tunneling from the half-space in
which the defect is situated) the $\psi _{\mathbf{k}1}^{\left( -\right)
}\left( \mathbf{r}\right) $ is written as%
\begin{equation}
\psi _{\mathbf{k}1}^{\left( -\right) }\left( \mathbf{r}\right) =\psi _{%
\mathbf{k}1}^{\left( 0\right) }\left( \mathbf{r}\right) +  \label{psi12} \\
\frac{im^{\ast }k^{3}a^{2}zz_{0}}{\hbar ^{2}rr_{0}}T(\widetilde{k}) t(%
\widetilde{k}) \psi _{\widetilde{\mathbf{k}}0}^{\left( 0\right) }\left(
\mathbf{r}_{0}\right) h_{1}^{\left( 1\right) }\left( kr\right) h_{1}^{\left(
1\right) }(\widetilde{k}r_{0}) .  \nonumber
\end{equation}%
Here $\psi _{\mathbf{k}0\mathbf{,}1}^{\left( 0\right) }\left( \mathbf{r}%
\right) $ and $T\left( k\right) $ are given by Eqs. (\ref{psi00}), (\ref%
{ps01}) and (\ref{T_d}). The wave functions (\ref{psi111}) and (\ref{psi12})
have a completely different form: In the lower half-space the wave function (%
\ref{psi111}) is the superposition of the transmitted p-wave $\psi _{\mathbf{%
k}1}^{\left( 0\right) }\sim $ $h_{1}^{\left( 1\right) }\left( kr\right) $, (%
\ref{ps01}), and two s-waves, one of which, $h_{0}^{\left( 1\right) }\left(
k\left\vert \mathbf{r}-\mathbf{r}_{0}\right\vert \right) $, is the wave
scattered by the defect and other one $h_{0}^{\left( 1\right) }\left(
k\left\vert \mathbf{r}-\widetilde{\mathbf{r}}_{0}\right\vert \right) $ is
the scattered wave, which undergoes reflection from the interface at $z=0$
(the wave moving from the 'image' defect placed in the mirror point $%
\widetilde{\mathbf{r}}_{0},$ $\left\vert \mathbf{r}_{0}-\widetilde{\mathbf{r}
}_{0}\right\vert =2z_{0}$). In the upper half-space there is only the p-wave
$\psi _{\mathbf{k}1}^{\left( -\right) }\sim h_{1}^{\left( 1\right) }\left(
kr\right) $, the amplitude of which depends on the scattering on the defect
because the wave incident to the contact is not a plane wave in this case.

\section{Total current and conductance}

The tunneling current $I( V) =I^{\left( +\right) }( V) -I^{\left( -\right)
}( V) $ is the difference between two currents flowing through the contact
in opposite directions. Each of them can be evaluated \ by means of the
probability current density integrated over a half-sphere of arbitrary
radius $r$, centered at the point contact $r=0$ and covering the contact
from the appropriate side, and integrating over all directions of the
electron wave vector. In this case the integrated probability current
density $J_{k}^{\left( \pm \right) }( V) $ is written as
\begin{eqnarray}
J_{k}^{\left( \pm \right) }( V) &=&-\frac{r^{2}\hbar }{m^{\ast }}\int
d\Omega \Theta \left( \pm z\right) \int d\Omega _{\mathbf{k}}\Theta \left(
\pm k_{z}\right)  \nonumber \\
&&Im\left( \psi _{\mathbf{k}1}^{\left( \pm \right) }( \mathbf{r}) \frac{%
\partial \psi _{\mathbf{k}1}^{\left( \pm \right) \ast }( \mathbf{r}) }{%
\partial r}\right) ,  \label{Ik}
\end{eqnarray}
where $d\Omega $ and $d\Omega _{\mathbf{k}}$ are elements of solid angle in
the real and momentum spaces, respectively. The total current through the
contact is%
\begin{equation}
\eqalign{I( V) =\frac{2e}{\left( 2\pi \right) ^{3}}\int\limits_{0}^{\infty }dkk^{2}%
\left[ J_{k}^{\left( +\right) }( V) f_{\mathrm{F}}\left( \varepsilon _{%
\mathbf{k}}-eV\right) \times \right.  \label{I} \\
\left( 1-f_{\mathrm{F}}\left( \varepsilon _{\mathbf{k}}\right) \right)
-\left. J_{k}^{\left( -\right) }(V) f_{\mathrm{F}}\left( \varepsilon _{%
\mathbf{k}}\right) \left( 1-f_{\mathrm{F}}\left( \varepsilon _{\mathbf{k}%
}-eV\right) \right) \right] ,}
\end{equation}%
where $f_{\mathrm{F}}\left( \varepsilon _{\mathbf{k}}\right) $ is the Fermi
function. At zero temperature only one of the terms in square brackets in
Eq.~(\ref{I}) differs from zero, i.e. only in one of the half-spaces states
are available for tunneling, depending on the sign of the bias. Using the
wave functions (\ref{psi111}) and (\ref{psi12}), after integration through
Eq.~(\ref{Ik}) the electrical current $I^{\left( \pm \right) }(V) $ at $|eV|
< \varepsilon_{\mathrm{F}}$ and $T=0$ takes the form

\begin{equation}
I^{\left( \pm \right) }( V) =\frac{e\hbar a^{4}}{36\pi m^{\ast }}
\times \label{IV}
\int\limits_{k_{\mathrm{F}}}^{\sqrt{k_{\mathrm{F}}^{2}+2m^{\ast
}\left\vert eV\right\vert /\hbar ^{2}}}dkk^{5}\left\vert t(
\widetilde{k}) \right\vert ^{2} \left( 1+\Phi( k^{\left( \pm
\right) }) \right) ,  \nonumber
\end{equation}%
where the integration is carried out over the absolute value of the wave
vector $k$ within the interval $\left\vert eV\right\vert $ of allowed
energies. We define $k^{\left( +\right) }=k$, $k^{\left( -\right) }=$ $%
\widetilde{k}=\sqrt{k^{2}-2m^{\ast }\left\vert eV\right\vert /\hbar ^{2}}$, $%
k_{\mathrm{F}}$ is the Fermi wave vector,
\begin{equation}
\eqalign{\Phi \left( k\right) =D^{-1}\sin \delta
_{0}\frac{z_{0}^{2}}{r_{0}^{2}}\left[ 12j_{1}\left( kr_{0}\right)
\right. \left( -y_{1}\left( kr_{0}\right) \cos
\delta _{0}\right. +  \nonumber \\
\left. \left\{ j_{1}\left( kr_{0}\right) \left( j_{0}\left( 2kz_{0}\right)
-1\right) +y_{0}\left( 2kz_{0}\right) y_{1}\left( kr_{0}\right) \right\}
\sin \delta _{0}\right) +  \nonumber \\
\left. 6\left( 1-j_{0}\left( 2kz_{0}\right) \right) \left(
kr_{0}\right) ^{-4}\left( 1+\left( kr_{0}\right) ^{2}\right) \sin
\delta _{0}\right] ,}
\end{equation}
and
\begin{equation}
D=1+2\sin \delta _{0} \times \\
\left[ \left( \frac{1}{2\left( 2kz_{0}\right) ^{2}} -j_{0}( 2kz_{0}) \right)
\sin \delta _{0} - y_{0}( 2kz_{0}) \cos \delta _{0} \right],  \nonumber
\end{equation}
and $j_{l}( x) $ and $y_{l}( x) $ are the spherical Bessel functions. From
Eq.~(\ref{IV}) it follows that the current-voltage dependence need not be
symmetric in voltage in the presence of a defect.

The differential conductance $G=dI/dV$ for $\left\vert eV\right\vert <
\varepsilon _{\mathrm{F}} $ and for $eV>0$ is, given by
\begin{equation}
G(V) =G_{0}\left[ q(V) \left( 1+\Phi(\widetilde{k}_{\mathrm{F}}) \right) -%
\frac{2}{k_{\mathrm{F}}^{4}} \int\limits_{k_{\mathrm{F}}}^{\widetilde{k}_{%
\mathrm{F}}}dkk^{5}\Phi \left( k\right) \right] ,  \label{G>}
\end{equation}%
and for $eV<0$,
\begin{equation}
G( V) =G_{0}\left[ q(V) +\frac{\widetilde{k}_{\mathrm{F}}^{2}}{k_{\mathrm{F}%
}^{2}}\Phi ( \widetilde{k}_{\mathrm{F}}) -\frac{4}{k_{\mathrm{F}}^{4}}%
\int\limits_{k_{\mathrm{F}}}^{\widetilde{k}_{\mathrm{F}}}dkk^{3}\widetilde{k}%
^{2}\Phi \left( k\right) \right] .  \label{G<}
\end{equation}%
Here $\widetilde{k}_{\mathrm{F}}=\sqrt{k_{\mathrm{F}}^{2}+2m^{\ast }eV/\hbar
^{2}}$ and,
\begin{equation}
q(V) =1+\frac{2m^{\ast }\left\vert eV\right\vert }{\hbar ^{2}k_{\mathrm{F}%
}^{2}}-\frac{1}{3}\left( \frac{2m^{\ast }\left\vert eV\right\vert }{\hbar
^{2}k_{\mathrm{F}}^{2}}\right) ^{3}.
\end{equation}%
\begin{equation}
G_{0}=\left\vert t\left( k_{\mathrm{F}}\right) \right\vert ^{2}\frac{%
e^{2}\left( k_{\mathrm{F}}a\right) ^{4}}{36\pi \hbar }  \label{G00}
\end{equation}%
is the conductance of the tunnel point-contact in the absence of a defect in
the limit $V\rightarrow 0$. At low voltage the conductance can be expressed
as an expansion in the parameter $1/\left(k_{\mathrm{F}}z_{0}\right)<1$,
\begin{equation}
\eqalign{G(0) = G_{0}\left\{ 1+12\frac{z_{0}^{2}}{r_{0}^{2}} \frac{1}{\left( k_{%
\mathrm{F}}r_{0}\right)^{2}} \sum\limits_{n=1}^{\infty }\left(
-1\right)^{n} \frac{\sin^{n}\delta_{0}}
{\left(2k_{\mathrm{F}}z_{0}\right)^{n-1}} \times
\right.  \nonumber \\
\left[ \frac{1}{2}\left( 1-\frac{1}{\left( k_{\mathrm{F}}r_{0}\right) ^{2}}%
\right) \sin \left( 2k_{\mathrm{F}}\left( r_{0}+\left( n-1\right)
z_{0}\right) +n\delta _{0}\right) +\right.  \nonumber \\
\left. \left. \frac{1}{k_{\mathrm{F}}r_{0}}\cos \left(
2k_{\mathrm{F}}\left( r_{0}+\left( n-1\right) z_{0}\right)
+n\delta _{0}\right) \right] \right\} \label{G(0)}}
\end{equation}
The second term in the Eq.~(\ref{G(0)}) corresponds to the sum over $n$
scattering events by the defect and $n-1$ reflections by the surface. If we
keep only the term for $n=1$ Eq.~(\ref{G(0)}) is consistent with the results
obtained by perturbation theory previously \cite{Avotina1,Avotina2,Avotina3}.

\section{Discussion and application to Kondo scattering}

The expansion (\ref{G(0)}) of the conductance $G$ demonstrates that as a
result of multiple scattering the conductance $G_{0}$, Eq.~(\ref{G00}), of
the tunnel point contact becomes modified with oscillatory contributions $%
\Delta G_{n}$, which at $1/\left( k_{ \mathrm{F}}z_{0}\right) \ll 1$ and $%
z_{0}\simeq r_{0}$ is of order
\begin{equation}
\Delta G_{n}\sim \frac{1}{\left( k_{\mathrm{F}}r_{0}\right) ^{n+1}}\sin
\left( 2k_{\mathrm{F}}\left( r_{0}+\left( n-1\right) z_{0}\right) +n\delta
_{0}\right) ,  \label{G_n}
\end{equation}%
where $n=1,2...$ is the number of scattering events on the defect placed at
a distance $r_{0}$ from the contact (and at a distance $z_{0}$ from the
interface), and $\left( n-1\right) $ is the number of reflections by the
interface. The argument of the sine function in Eq.~(\ref{G_n}) corresponds
to the phase the electron accumulates while moving along a semiclassical
trajectory. In Fig.~\ref{fig1}(a,b) such trajectories are illustrated for
the case of scattering twice by the defect and one specular reflection by
the interface. For $eV>0$ (Fig.\ref{fig1}a) this trajectory consists of a
segment (labelled 2) passing through the contact and arriving at the defect,
two line segments (3 and 4) connecting the defect and the interface (these
segments are perpendicular to the interface because only along such
trajectory the electron can return to the defect and undergo the second
scattering), and the part (5) from the defect to the contact. After specular
reflection from the contact this wave interferes with the partial wave (1)
that is directly transmitted through the contact.
\begin{figure}[tbp]
\includegraphics [width=8cm]{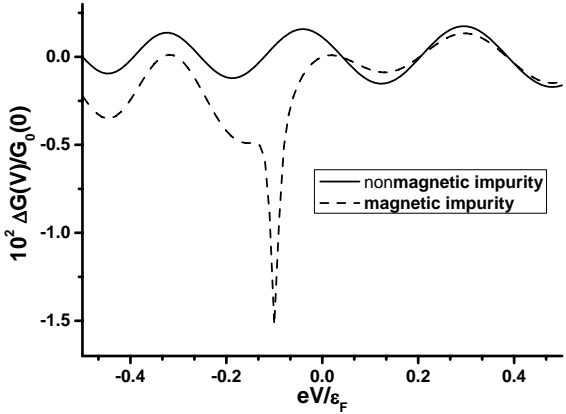}
\caption{Voltage bias dependences of the normalized conductance corrections $%
\Delta G(V) /G_{0}$ for a magnetic and a nonmagnetic impurity calculated
from Eqs.~(\protect\ref{G>}) and (\protect\ref{G<}). We have used the
parameters $\protect\varepsilon _{\mathrm{K}}=0.9\protect\varepsilon _{%
\mathrm{F}},$ $k_{\mathrm{B}}T_{\mathrm{K}}=0.01\protect\varepsilon _{%
\mathrm{F}},$ $r_{\mathrm{D}}=0.1\protect\lambda _{\mathrm{F}}/2\protect\pi %
, $ $\protect\rho _{0}=0,$ and $z_{0}=20\protect\lambda _{\mathrm{F}}/2%
\protect\pi .$}
\label{fig2}
\end{figure}

When $eV<0$ (Fig. \ref{fig1}b) a wave incident to the contact (trajectory 2)
is partially reflected from the contact. The electron moving along the
trajectory 3 from the contact to the defect is partially scattered towards
the interface (line segments 4) where it undergoes specular reflection from
the interface (5) and comes back to the defect, from which it returns to the
contact via trajectory 6. Tunnelling through the contact this partial wave
interferes with the partial wave that is directly transmitted (1) in the
half-space $z<0$. At each scattering on the defect the electron acquires an
additional phase shift $\delta _{0}.$ The phase shift $\Delta \phi $ between
the two interfering partial waves for an electron with wave vector $\mathbf{k%
}$ is $\Delta \phi =2kr_{0}+2kz_{0} +2\delta _{0}.$ Because the maximum
value of the electron wave vector depends on the applied voltage the
conductance oscillates as the function of $eV$.

\{The differential conductance, as the derivative of the current,
discriminates a bound of the energy interval, which depends on the bias $eV,$
i.e. for $eV>0$ the period of oscillations is defined by the energy $%
\varepsilon _{\mathrm{F} }+eV$ and for $eV<0$ - by the energy $\varepsilon _{%
\mathrm{F}}-\left\vert eV\right\vert .$

However, the current voltage characteristics is not symmetric relative to
the point $V=0$. This asymmetry results from the dependencies of the phase
shift $\delta _{0}(\widetilde{k}_{\mathrm{F}})$ and the absolute value of
the wave vector $\widetilde{k}_{\mathrm{\ F}}=\sqrt{k_{\mathrm{F}%
}^{2}+2m^{\ast }eV/\hbar ^{2}}$ on the sign of $eV $. The physical origin of
this asymmetry is that the scattering depends on the electron energy in the
lower half-space, which is different for different directions of the current.

The dependence $\delta _{0}(k) $ on $k$ is defined by the form of the
scattering potential $U(r)$. To illustrate the obtained results for an
s-wave phase shift we use the following model function \cite{corral,Shift},
\begin{equation}
\delta _{0}(k) =\delta _{0\mathrm{K}}+\delta _{0\mathrm{D}}= \left[ \frac{%
\pi }{2}-\tan ^{-1}\left( \frac{\varepsilon _{\mathbf{k} }-\varepsilon _{%
\mathrm{K}}}{T_{\mathrm{K}}}\right) \right] -kr_{\mathrm{D}}.  \label{d0}
\end{equation}%
The first term in Eq.~(\ref{d0}) describes the resonant scattering on a
Kondo impurity level $\varepsilon _{\mathrm{K}}$ ($T_{\mathrm{K}}$ is the
Kondo temperature). For $\varepsilon _{\mathbf{k}}\rightarrow \varepsilon _{%
\mathrm{K}}$ the effective electron scattering cross section acquires a
maximum value corresponding to the Kondo phase shift $\delta _{0\mathrm{K}%
}=\pi /2$ \cite{Abrikosov}. For a non-magnetic impurity this term
is absent. The second term takes into account the usual potential
scattering. For simplicity we use the s-wave phase shift for a
hard sphere potential of radius $r_{ \mathrm{D}}$
($k_{\mathrm{F}}r_{\mathrm{D}}<1$). Inelastic scattering by the
magnetic defect can be taken into account in the scattering
formalism by introducing an imaginary part of the phase
(\ref{d0}).

Figure~\ref{fig2} shows the dependences of the corrections to the normalized
conductance $\Delta G(V) /G_{0}=$ $\left( G(V) -G_{0}(V) \right) /G_{0}$
resulting from the scattering by a defect placed on the contact axis for a
magnetic and a nonmagnetic impurity. The figure illustrates the appearance
of a Kondo anomaly in the conductance seen as an extremum in the
differential conductance, $G(V) $, near the bias $eV_{\mathrm{K}}$
corresponding to the resonance condition $\varepsilon _{\mathrm{F}}+eV_{%
\mathrm{K}}-\varepsilon _{ \mathrm{K}}=0$. The plots show a slowly
increasing background on top of the oscillating $\Delta G(V) $ dependence.
The background arises from the integral terms in Eqs.~(\ref{G>}), (\ref{G<}%
), which take into account the contribution of all available states within
interval $\left\vert eV\right\vert $. The monotonic part in $\Delta G(V) $
is more pronounced in the case of Kondo scattering, which gives a large
contribution to this part at any voltage.

It is interesting to observe that the sign of the Kondo anomaly depends on
the distance between the contact and the defect $r_{0}$. This distance in
combination with the value of the wave vector $\widetilde{k}_{\mathrm{F}}$
determines the period of oscillation of $\Delta G(V) $, which is indeed a
non-monotonic function of $\widetilde{k}_{ \mathrm{F}}r_{0}$. If the bias $%
eV_{\mathrm{K}}$ coincides with a maximum in the oscillatory part of
conductance the sign of the Kondo anomaly is positive and vice versa, the
negative sign of the Kondo anomaly is found at a minimum in the periodic
variation of $\Delta G.$

\begin{figure}[tbp]
\includegraphics[width=8cm]{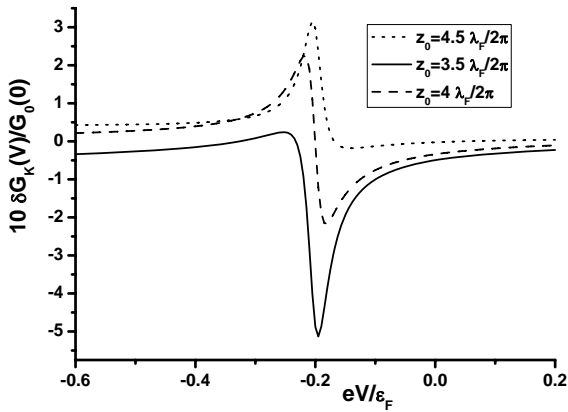}
\caption{Difference $\protect\delta G_{\mathrm{K}}(V) /G_{0}$ between the
voltage dependences of the conductance for a magnetic and a non-magnetic
impurity. We have used the parameters $\protect\varepsilon _{\mathrm{K}}=0.9%
\protect\varepsilon _{\mathrm{F}},$ $k_{\mathrm{B}}T_{\mathrm{K}}=0.01%
\protect\varepsilon _{\mathrm{F}},$ and $r_{\mathrm{D}}=0.1\protect\lambda _{%
\mathrm{F} }/2\protect\pi .$}
\label{fig3}
\end{figure}

\begin{figure}[tbp]
\includegraphics [width=8cm]{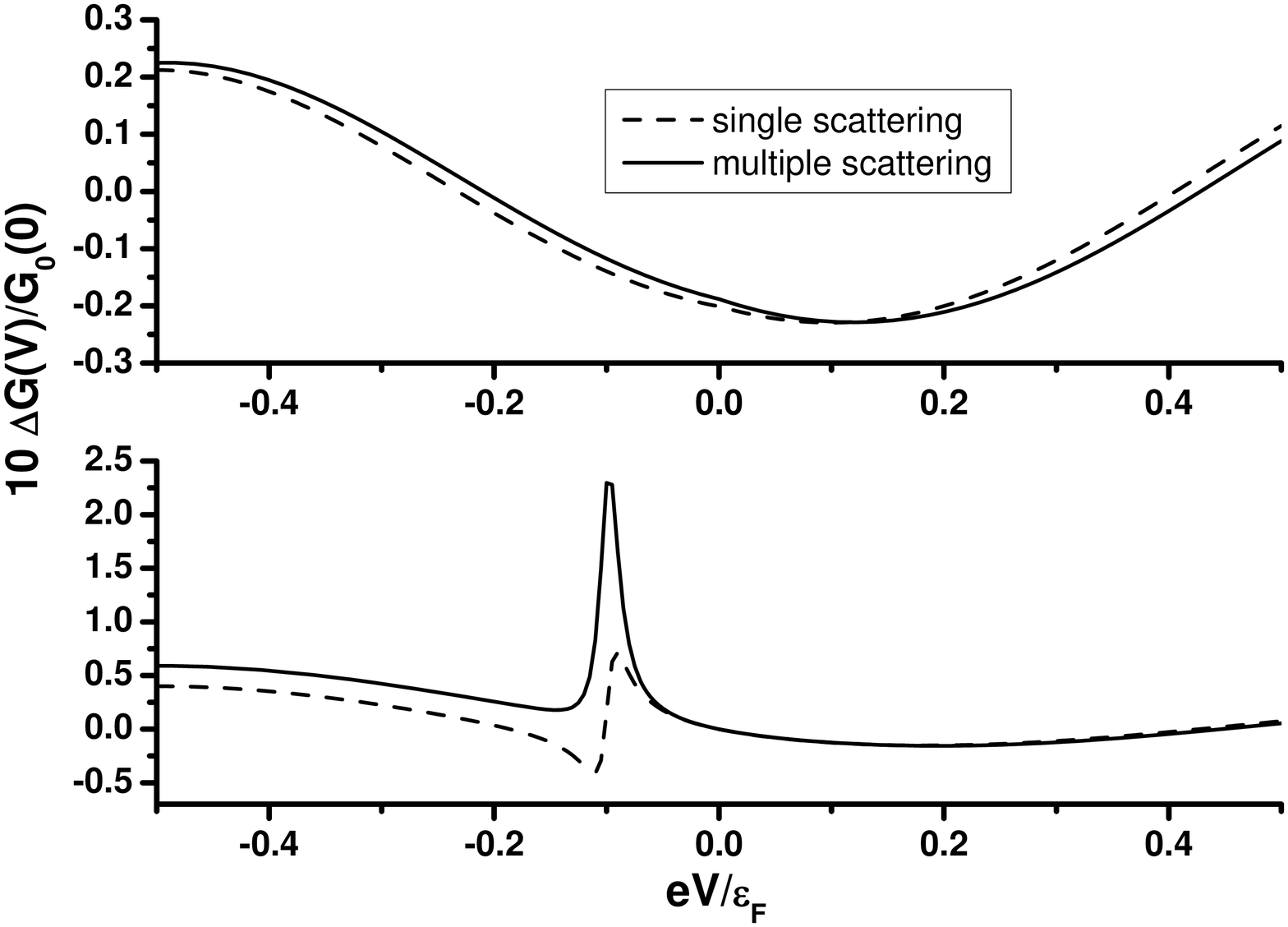}
\caption{Comparison of the oscillatory parts of the conductance $\Delta G(V)
/G_{0}$ calculated by using the Eqs.~(\protect\ref{G>}, \protect\ref{G<})
(full curves) and by means of results obtained in the framework of
perturbation in the electron-impurity interaction (dashed curves). $a$ -
non-magnetic defect; $b$ - magnetic defect. We have used the parameters $%
\protect\varepsilon _{\mathrm{K}}=0.9\protect\varepsilon _{\mathrm{F}},$ $k_{%
\mathrm{B}}T_{\mathrm{K}}=0.01\protect\varepsilon _{\mathrm{F}},$ $r_{%
\mathrm{D}}=0.1\protect\lambda _{\mathrm{F}}/2\protect\pi ,$ $\protect\rho %
_{0}=0,$ and $z_{0}=5\protect\lambda _{\mathrm{F}}/2\protect\pi .$}
\label{fig4}
\end{figure}
In Fig.\ref{fig3} we present the difference $\delta
G_{\mathrm{K}}(V) /G_{0}=(\Delta G_{m}-\Delta G_{n})/G_{0}$
between voltage dependences for a magnetic $\Delta G_{m}$ and a
non-magnetic $\Delta G_{n}$ impurity, having the same potential
scattering strength. The plots in the Fig.~\ref{fig3} show the
evolution of the shape of the Kondo anomaly for several values of
the distance between the contact and the impurity, placed on the
contact axis. The change of distance changes the periodicity of
the normal-scattering oscillations which is illustrated to lead to
a changing of sign in the Kondo signal. A similar dependence of
the differential conductance with the distance between an STM tip
and an adatom on the surface of a metal has been obtained
theoretically in Refs.~\cite{Zawad,Lin} in the terms of the
Anderson impurity Hamiltonian \cite{Anderson}. Note that we
obtained the Fano-like shape of the Kondo resonance in the
framework a single-electron approximation while in
Refs.~\cite{Zawad,Lin} the many-body effects were taken into
account. 

Figure~\ref{fig4} illustrates the importance of multiple scattering for this
problem. It shows the oscillatory parts of the conductance $\Delta G(V)
/G_{0}$ calculated by using the Eqs.~(\ref{G>}, \ref{G<}) in comparison to
results obtained in the framework of perturbation in the electron-impurity
interaction \cite{Avotina1,Avotina2}, i.e. neglecting multiple electron
scattering. While for the non-magnetic impurity ((Fig.~\ref{fig4}a)) the
difference between two curves is small it is seen that for a magnetic
impurity (Fig.~\ref{fig4}b) the perturbation method does not describe the
conductance correctly in a region of the Kondo resonance. For nonmagnetic
impurities multiple scattering has a negligible effect due to the smallness
of contributions of the multiple scattering paths described by the the
parameter $\left( k_{\mathrm{\ F}}z_{0}\right) ^{-1},$ which is no longer
true near the Kondo resonance, where the increasing of the scattering
amplitude is the dominant effect.

\section{Conclusion}

We have studied the influence of multiple electron scattering by a single
defect on the current through a tunnel point-contact. In the approximation
of s-wave scattering by the defect a general expression for the conductance $%
G$ has been found (\ref{G>}), (\ref{G<}). The results obtained have been
analyzed for the model s-wave phase shift (\ref{d0}) describing the Kondo
scattering by a magnetic impurity. We demonstrated that taking multiple
scattering into account is most essential near voltage values corresponding
to the Kondo resonance condition $\varepsilon _{ \mathrm{F}}+eV=\varepsilon
_{\mathrm{K}}$. It is found that the the shape as well as the sign of the
Kondo anomaly depends on the position of the defect. This dependence results
from quantum interference of partial waves directly transmitted through the
contact with the partial wave scattered by the defect and reflected by the
interface. The phase shift between the two waves produces the oscillations
of the conductance. A maximum in the regular oscillation of $G$ leads to a
positive sign of the Kondo anomaly at that position, while a minimum
produces a negative sign. These results may be exploited in future
experiments for detecting and investigating the Kondo effect of individual
impurities in the bulk of a host metal.

\ack Ye. S. A. is supported by the INTAS grant for Young
Scientists (No 04-83-3750) and Yu. A. K. was supported by a NWO
visitor's grant. This research was supported supported partly by
the program "Nanosystems nanomaterials, and nanotechnology" of
national Academy of Sciences of Ukraine.
\

\end{document}